\documentclass[preprint,3p,twocolumn]{elsarticle}

\usepackage[normalem]{ulem}
\usepackage{datetime}
\usepackage[utf8]{inputenc}
\usepackage[T1]{fontenc}
\usepackage{subcaption}
\usepackage{amsmath}
\usepackage{tabularx}
\usepackage{booktabs}
\usepackage{minted}
\usepackage{lipsum}

\graphicspath{{images/}}

\newcolumntype{R}{>{\raggedleft\arraybackslash}X}

\usepackage{listings}
\DeclareRobustCommand*\chem[1]
 {\ensuremath{%
   {\mathcode`\-="0200\mathcode`\=="003D
    \ifx\f@series\testbx\mathbf{#1}\else\mathrm{#1}\fi}}}

\newif\ifdraft
\drafttrue
\ifdraft
  \usepackage[dvipsnames]{xcolor}
  \usepackage{soul}
  \usepackage{xcolor}
  \usepackage{etoolbox} 
  \usepackage{tcolorbox}
\newcommand{\bsnote}[1]{\textcolor{red}{[BS: #1]}}
\newcommand{\msnote}[1]{\textcolor{blue}{[MS: #1]}}
\newtcolorbox{msnotes}{
  colback=white,
  colframe=blue,
  fontupper=\color{blue},
  boxrule=0.5pt,
  arc=0mm,
  boxsep=1pt,
  left=1pt,
  right=1pt,
  top=1pt,
  bottom=1pt,
  before upper={\textcolor{blue}{MS:\\ }}, 
}
\newtcolorbox{todos}{
  colback=white,
  colframe=magenta,
  fontupper=\color{magenta},
  boxrule=0.5pt,
  arc=0mm,
  boxsep=1pt,
  left=1pt,
  right=1pt,
  top=1pt,
  bottom=1pt,
  before upper={\textcolor{blue}{Todo:\\ }}, 
}
\newcommand{\todo}[1]{\textcolor{magenta}{[TODO: #1]}}
\else 
\newcommand{\bsnote}[1]{}
\newcommand{\msnote}[1]{}
\newcommand{\todo}[1]{}
\fi

\newif\ifshoworiginal
\showoriginaltrue
\ifshoworiginal
\newcommand{\soriginal}[1]{\sout{#1}}
\else
\newcommand{\soriginal}[1]{}
\fi

\usepackage{hyperref}
\hypersetup{
    colorlinks=true,
    linkcolor=blue,
    citecolor=blue,      
}

\begin{document}

\begin{frontmatter}
  \title{XenoFlow: How Fast Can a SmartNIC-Based DNS Load Balancer Run? \\
    {\large An Evaluation of DOCA Flow on the Nvidia BlueField-3}}

  \author[1]{Max Schr\"otter} 
  \ead{max.schroetter@uni-potsdam.de}

  \author[1]{Sten Heimbrodt}
  \ead{sten.heimbrodt@uni-potsdam.de}

  \author[1]{Bettina Schnor}
  \ead{schnor@cs.uni-potsdam.de}

  \affiliation[1]{organization={University of Potsdam},
    addressline={An der Bahn 2},
    postcode={14476},
    city={Potsdam},
    country={Germany}}

  \begin{abstract}
    With the advent of programmable network hardware, more and more
    functionality can be moved from software running on general purpose CPUs to the
    NIC\@. Early NICs only allowed offloading fixed functions like checksum
    computation. Recent NICs like the Nvidia Bluefield-3 allow a fully programmable
    dataplane.
    In this paper, we present our first steps 
    towards a load balancer named XenoFlow running on the Bluefield-3. Furthermore, we show the
    capabilities and limitations of the Bluefield-3 eSwitch. Our results show
    that 
    the Bluefield-3 will not achieve line rate with only 2 entries in a Flow
    Pipe. However, we also show the adventages of hardware offloading on the NIC and being
    closer to the network. With XenoFlow, we achieve an 44\%~lower latency
    compared to a comparable eBPF-based load balancer running on the host.
    Furthermore, XenoFlow achieves this low latency even under high load.
  \end{abstract}

\end{frontmatter}

\section{Introduction}\label{sec:intro}

Load balancers are a core component in modern data centers. They distribute the
traffic to multiple servers, which allows scaling services beyond the
capacity of a single node. Load balancers can distribute traffic based on
information of different layers of the OSI model. For this work, we focus on
Layer 3/Layer 4 load balancers, which distribute traffic based on the IP and
TCP/UDP headers. Since most services store stateful information,
per-connection consistency (PCC)~\cite{PCC} is a core requirement of load
balancers.

Historically, L4 load balancers were implemented in software in user
space~\cite{HAProxy} or kernel space~\cite{IPVS}. Since the general purpose Linux
network stack limits the performance of load balancers, previous work has moved
load balancers into user space with bypassing the Linux network stack using
DPDK~\cite{DPDK} or similar~\cite{DPVS,maglev}. With the
introduction of eBPF \& XDP~\cite{xdp}, it became possible to implement L4 load
balancers before packets are processed by the Linux network stack.
Katran~\cite{katran} is a well-known example of such an eBPF-based L4 load
balancer, which is implemented in a few hundred lines of C code inside the linux
kernel eBPF virtual machine. While this approach is very efficient, it still
uses the CPU for processing packets and copies packets up to
main memory and back to the NIC\@.

To achieve higher throughput, the SilkRoad~\cite{SilkRoad}
  project moved a L4 load balancer closer to the network and
  offloaded the load balancer to a programmable switch in the network
path. The authors demonstrate that this still requires software components in the data
center to achieve per-connection consistency. 

This paper investigates first steps towards a L3 load balancer called XenoFlow
running on the Nvidia Bluefield-3 SmartNIC\@. While there has been work
implementing L7 and L4 load balancers on the
Bluefield-3~\cite{LBSNIC21,COMPBF,laconic} to our knowledge, all of them still
run partially on the general purpose ARM cores of the Bluefield-3 limiting the
performance.



Nvidia advertises the Bluefield-3 as 400 Gbit/s \textit{infrastructure compute
platform} with its NIC-Subsystem (DPA + eSwitch) accelerating
data path at line rate~\cite{HC33_NV_MARKETING,NV_BF_MARKETING}.
While Chen et al.~\cite{DEMYSTIFYINGBF3} have already
shown that the Bluefield-3 DPA
will not achieve line rate for just sending or receiving 64 byte packets, this
has not been investigated for the eSwitch.

Here we present the results of our first steps on the Bluefield-3 which answer the
following research question:
\begin{enumerate}
\item {\bf RQ1:} Can we implement a simple load balancer using the Bluefield
  DOCA Flow API\@?
\item {\bf RQ2:} Do DOCA Flow applications process with zero overhead, i.e.\ do no packets
  get dropped below the maximum throughput?
\item {\bf RQ3:} Can the BlueField-3 achieve line rate with a load balancer
  implemented using the DOCA Flow API\@?
\item {\bf RQ4:} How does our load balancer affect the end to end latency?
\item \textbf{RQ5:} How does load effect the performance of the load balancer?
\end{enumerate}

The paper is organized as follows: First we  review related work, followed by
a description of the relevant hardware components and the software stack of
the Bluefield-3. Then we present our design and implementation of
\emph{XenoFlow} in Section~\ref{sec:pht} and show our preliminary performance results
in Section~\ref{sec:evaluation}. Finally, we discuss the open challenges and future work.

\section{Related Work}%
\label{sec:related}


This section discusses related work regarding the Bluefield SmartNIC\@.


Liu et al.~\cite{PERFBF2} have compared the performance of the
BlueField-2 ARM cores against 12 different x86 and 2 ARM servers. All servers
had similar core counts and clock speeds. They found that the performance
capability of the BlueField-2 lacks behind the x86 servers, except for workloads
the BlueField-2 has accelerators for, like e.g.\ the linux cryptography API
\texttt{af-alg}. Furthermore, they found that the ARM cores could not saturate the
bandwith of the 100 Gbit/s NIC using \texttt{pktgen}.

Michalowicz et al.~\cite{COMPBF} have compared the Bluefield-2 and the
Bluefield-3 from the HPC perspective. 
The authors measure the inter DPU latency and bandwidth running the
OSU-Benchmarks utilizing the MVAPICH MPI library on the DPUs.
The inter-DPU benchmarks show an up to 1.78$\times$ better latency and
1.5$\times$ better network bandwidth of the Bluefield-3 compared to the Bluefield-2. The
authors attribute this to the increased bandwidth of the Bluefield 3.
However, the reached bandwidth is below 10 GB/s even for 16KiB
messages.

In the context of applications with a DPU-aware MPI library, the authors showed
minor improvements for a modified P3DFFT application
compared to the host system. 

With the introduction of the Bluefield-3, Nvidia introduced a datapath
accelerator (DPA) which they characterize as ``having many execution units that
can work in parallel to overcome latency issues (such as access to host memory)
and provide an overall higher throughput''~\cite{NV_BF_MARKETING}.
However, Chen et al.~\cite{DEMYSTIFYINGBF3} have shown that the DPA cores are
much wimpier than the ARM and host cores. Furthermore, if packet size becomes
larger than 1KB the DPA must rely on the slower ARM/host memory to achieve line
rate. In the experiments, the DPA did not achieve the network throughput of the ARM or
host cores for any
packet size. However, in latency critical applications like clock synchronization
the DPA achieved a 2 x lower time uncertainty bound than the ARM/host cores.
Also, the Authors derive suggestions from their experiments how to improve the
DPA design: ``Directly attach a memory to the DPA'' and ``Equip DPA with a more
powerful cache''.

Cui et al.~\cite{laconic,laconic-early} present a L7 load balancer called Laconic for the
BlueField-2. The request parsing logic executed on
the ARM
cores of the DPU\@. Because the authors measured a rule insertion latency of up
to 305$\mu s$ they decided to offload only the rewriting of reply packets from large
flows to the eSwitch. Smaller flows are completely processed on the ARM cores limiting
the performance for small flows. For flows of 4~MB or more, Laconic
outperforms the x86-based nginx load balancer. This shows the power of the
Bluefield eSwitch for packet rewriting.



\section{The Bluefield-3}

Here we present the hardware features and the software API of the
Bluefield-3, especially the ones which are interesting for the
implementation of \emph{XenoFlow}.

\subsection{Hardware Features}

\begin{figure}
  \begin{center}
    \includegraphics[width=0.5\textwidth]{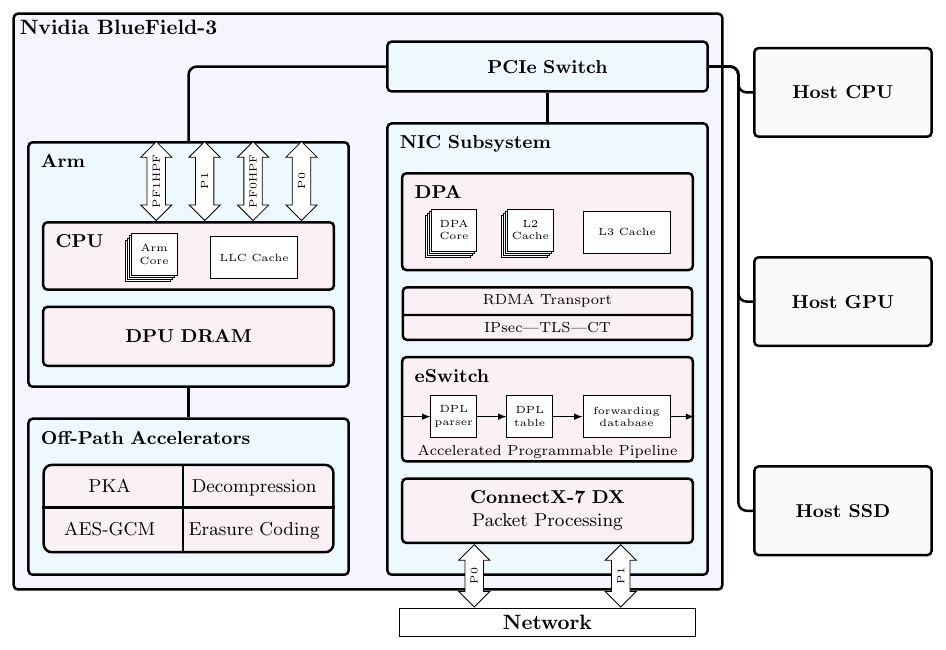}
    \caption{Architecture of the Nvidia Bluefield-3 DPU.}%
    \label{fig:bf3-arch}
  \end{center}
\end{figure}
The Bluefield-3 consists of the following components as shown in
Figure~\ref{fig:bf3-arch}:
\begin{enumerate}
  \item 16 ARM Cores running Linux (off-path),
  \item Off-Path Accelerators, sometimes also called lookaside accelerators:
    AES-GCM, Decompression, etc., 
  \item The Accelerated Programmable Pipeline (APP) consisting of 64--128 packet
    processing cores (on-path), and
  \item The Data-Path Accelerator (DPA) consisting of 16 hyperthreaded I/O and
    Packet Processing Cores~\cite{NV_BF_HWARC} (on-path). 
\end{enumerate}

While the APP and DPA cores are on-path, meaning they process every packet in
the default data path, the ARM cores and the lookaside accelerators are
off-path. They only process a packet if the data plane forwards the packet to
them. 

As described in Section~\ref{sec:related} there has been multiple work
deploying load balancers to the Bluefield-3 ARM cores and the DPA\@. This work
focuses on implementing a load balancer on the Accelerated Programmable
Pipeline using the DOCA Flow API (see Section~\ref{sec:software}).

While the Nvidia documentation specifies that APP consists of 64--128 packet processing
cores, there are no more details describing the kind of cores or their capabilities.
However, Nvidia's marketing material promises packet processing at line rate~\cite{HC33_NV_MARKETING,NV_BF_MARKETING}.
Furthermore, when looking at the physical Bluefield-3 hardware, only the dies for
the ARM cores and the DPA are visible. This makes it necessary to rely on
benchmarking to determine the capabilities of the packet processing cores.

Although the DOCA Flow API documentation never mentions the APP, it states
that the created rules are executed on the eSwitch. Since \emph{``modern network controllers have a
complex embedded switch''}~\cite{INTEL_ESWITCH} the eSwitch is probably the Accelerated
Programmable Pipeline (APP) consisting of the 64--128 packet processing cores.
An eSwitch is configured and controlled by the Physical Function driver (PF driver) per-port on the Ethernet controller~\cite{INTEL_ESWITCH}.

Furthermore, the DOCA Flow API never clearly states the capabilities
of the eSwitch as depicted in Figure~\ref{fig:bf3-arch}.
We assume that the limitations of
ASAP$^2$~\cite{MELLANOX_ASAP,NV_APAP_LIM} (e.g. 15 groups, 256K entries per
group), which is used for hardware offloading Open vSwitch (OVS) rules to the eSwitch, will
probably also apply to the DOCA Flow API\@. Rules and pipelines are offloaded to
the eSwitch via GRPC from the host. 

An overview of the compute components of the ARM, DPA and Host cores in our
system can be found in Table~\ref{tab:bf3-compute-components}.

\begin{table*}[t]
  \centering
  \caption{Compute Components of the Bluefield-3 DPU.}%
  \label{tab:bf3-compute-components}
  \begin{tabularx}{\textwidth}{lrrr}
    \toprule
    Resource   & Host & ARM & DPA \\
    \midrule
    CPU        & 2 $\times$ Intel Xeon Silver 4514Y & ARMv8.2+ A78 Hercules
    & RISC-V RV64IMAC(B)-USM\\
    Cores      & 2 $\times$ 16  & 16      & 16 \\
    Threads    & 2 $\times$ 32  & 32      & 256 \\
    Frequency  & 3.4~GHz        & 2.0~GHz\footnotemark& 1.8~GHz \\
    RAM        & 128~GB         & 32~GB   &  \\
    L1D Cache  & 2 $\times$ 16 $\times$ 48~KB  & 16 $\times$ 64~KB  & 256 $\times$ 1~KB \\
    L1I Cache  & 2 $\times$ 16 $\times$ 32~KB  & 16 $\times$ 64~KB  & 8~KB \\
    L2 Cache   & 2 $\times$ 16 $\times$ 2~MB   & 16 $\times$ 512~KB & 1.5~MB \\
    L3 Cache   & 2 $\times$ 30~MB   & 16~MB   & 3~MB \\
    \bottomrule
  \end{tabularx}
\end{table*}

\footnotetext{measured using \texttt{perf}}

\subsection{Software Stack for APP}\label{sec:software}

DOCA is a collection of libraries and APIs to access the hardware features of
the Bluefield ecosystem. In this ecosystem, the eSwitch is programmable via
multiple APIs:
\begin{itemize}
  \item DOCA Flow API: This is the main API to configure the eSwitch and
    offload rules to it. It is used to create pipelines, tables, and rules.
  \item OVS: The Open vSwitch (OVS) is a software switch that can be
    used to manage the eSwitch. It uses the DOCA Flow API to offload rules to
    the eSwitch.
\end{itemize}

For this work, we need a combination
of the DOCA Flow API and OVS to control the eSwitch.

The core building block of the DOCA Flow API are \textit{Pipes}.  A
Flow Pipe is a software template for a flow
table.\footnote{Introduction to Developing Applications with Nvidia
  DOCA on BlueField DPUs, 2022
  (\url{https://www.youtube.com/watch?v=H7T-yS4FTqI})} There is
one root pipe which acts as entry point for each packet. A
pipe specifies which packets to match on, which actions can be
performed on them, and where the packet can be forwarded. For
each pipe, multiple forwards can be defined if the packet matches, and
one if the packet does not match the \textit{match} filter.
Forwards can include forwarding packets to other pipes, allowing one to build a
tree of pipes. Furthermore, packets can be forwarded to the ARM cores
using Receive Side Scaling (RSS), to a different port, or can be dropped. 
However, a pipe is just a template for specifying the
  capabilities of its entries. Entries are the actual rules that are
offloaded to the eSwitch.  There are two ways to define
\textit{matches} and \textit{actions} of a pipe and its entries:
\begin{itemize}
  \item constant: The match criteria and actions are defined by the pipe
    itself and are constant for all entries of the pipe.
  \item variable: The match criteria and actions are defined by each entry of
    the pipe and can differ between entries. This requires the pipe match value
    to be set to all ones.
\end{itemize}
Furthermore, both can be defined
\begin{itemize}
  \item explicitly: The match criteria and actions define a mask,
    setting the bits that should be compared to or be changed, to 1. The match
    value then defines the values of those bits.
  \item implicitly: The mask is set to NULL and all bits of the field are
    compared or set to the value.
\end{itemize}

For example, if one wants to distribute packets to two backends based on the last
bit of the source IP address, the match needs to be defined as an explicit  
variable match. The pipe match value for the source IP address needs to be set
to 0xFFFFFFFF and the match mask to 0x00000001. While the first entry's value is
set to 0x00000000, the second entry's value is set to 0x00000001. This way 
the first entry matches all packets with an even source IP address and the
second entry matches all packets with an odd source IP address. The same concept
applies to actions.

During DOCA Flow API initialization, one of the following pipe
modes~\cite{NV_BF_PIPE}
can be selected: \emph{vnf, switch} or \emph{remote-vnf} mode.
These modes set the capabilities where packets can be forwarded to and predefine
the miss forward. 
In virtual network function (\emph{vnf}) mode 
packets arriving at one port are processed and can be sent out on another port.
Furthermore, packets missed by the \textit{match} filter are sent to
ARM\footnote{As we have seen in our experience}.
The \emph{switch} mode is
for internal switching and allows forwarding to uplink representor ports only. 
In the \emph{remote-vnf}
mode the DOCA Flow app is executed on the host. Packets missing the
filter are forwarded to the host which in turn can set up new rules via GRPC on
the Bluefield. 

For our use case we use the \emph{vnf} since we want to modify packets, change
its destination depending on the backend, and send it out on another port.

Packet modifications beyond the L4 headers, as well as arbitrary writes to these
headers, are not supported. All modifications must be statically defined within
a single action. However, since pipelines can be chained, more complex
modifications can be realized by storing intermediate results in packet
metadata. Moreover, DOCA Flow provides additional functionality, such as the
computation of checksums through specialized pipelines. If further processing is
required, the packet must be forwarded to the ARM or DPA cores for continued
processing.
\section{Design and Implementation of XenoFlow}\label{sec:pht}

In our approach implementing an L3 load balancer on the eSwitch, we
first focused on a basic prototype named XenoFlow. This prototype forwards packets by
rewriting the destination MAC addresses using DOCA Flow. Because of the limited
DOCA Flow documentation, we started by exploring the example applications
provided by Nvidia~\cite{NV_FLOW_EX}. All other undocumented features were
explored by trial and error.

\subsection{XenoFlow 0.1}
\begin{figure*}[t]
  \begin{center}
    \includegraphics[width=0.8\textwidth]{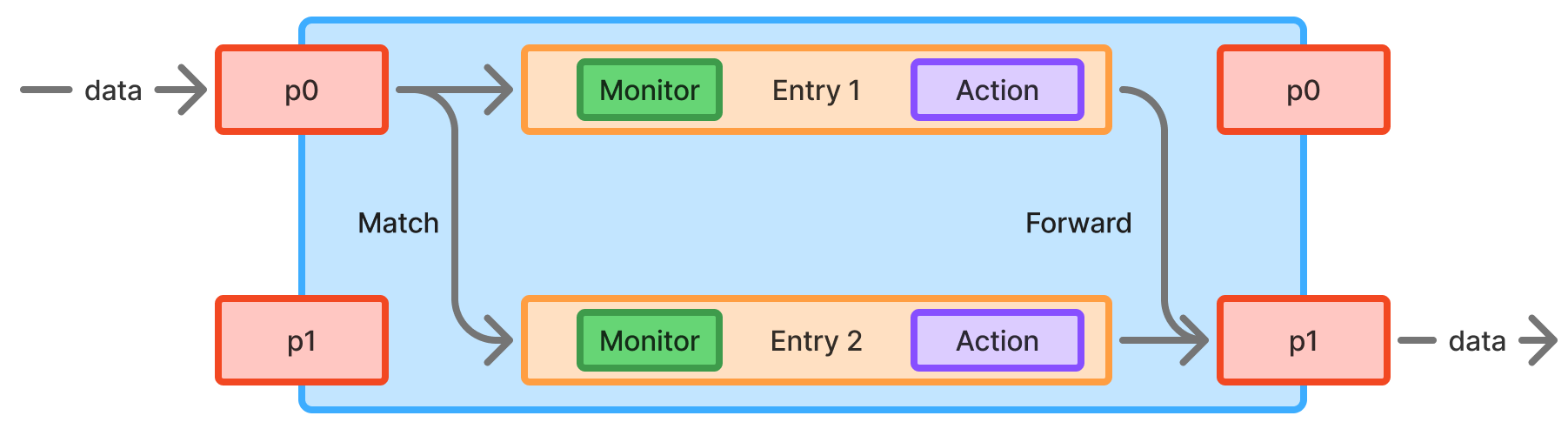}
    \caption{XenoFlow Root Pipe}%
    \label{fig:xenoflow-root-pipe}
  \end{center}
\end{figure*}

XenoFlow's main classification criterion is the IPv4 source address from
the header of the incoming IP packet. That source address is, if not modified,
the corresponding address of the actual sender. 
In our prototype, we assume that the IP address space exhibits sufficient
randomness, such that the source IP address can serve as a classification
criterion for partitioning flows into equally sized classes, which are
subsequently distributed across a predefined number of backends.
This classification criterion also corresponds with the capacities of the
eSwitch, which only allows matches and modifications on the fixed length parts
of a packet. This boils down to the fact that most hardware accelerators heavily
incorporate the usage of ASICs.

For our evaluation in Section~\ref{sec:evaluation} we used two backends. The
load balancing algorithm is therefore reduced to the decision whether the IPv4
address  is even or odd.

\subsection{Implementation with DOCA Flow}

The DOCA Flow library consists of header files and a closed source shared
library object. The header files declare the data structures which are used to
define rules that the actual hardware can apply to the incoming data. These data
structures are then applied using C functions that map to the corresponding
hardware functions of the ASIC\@.

XenoFlow uses a root pipe, which is by definition always the first pipe of an
application, that contains a set of entries that directly correspond to the
number of backends that are configured to process incoming traffic. The entries
get packets from the pipe according to their match rules, as can be seen in
Figure~\ref{fig:xenoflow-root-pipe}. These match rules evaluate the last bits of
each packet's IP source address field, using the implicit matching paradigm (see
Section~\ref{sec:software}). Each of these entries changes the destination MAC
address of the packet to the MAC address of the specific backend. Additionally,
the entry changes the hardware output port to the second BlueField connector. In
order to receive information about the current processing speed and the amount
of processed data by XenoFlow, we applied a monitor data structure to each
entry. These allow gathering information about the hardware's current
processing metrics. Two data points can be obtained:
\begin{itemize}
    \item number of processed packets
    \item amount of processed data in bytes
\end{itemize}
\begin{figure*}[t]
  \begin{center}
    \includegraphics[width=0.8\textwidth]{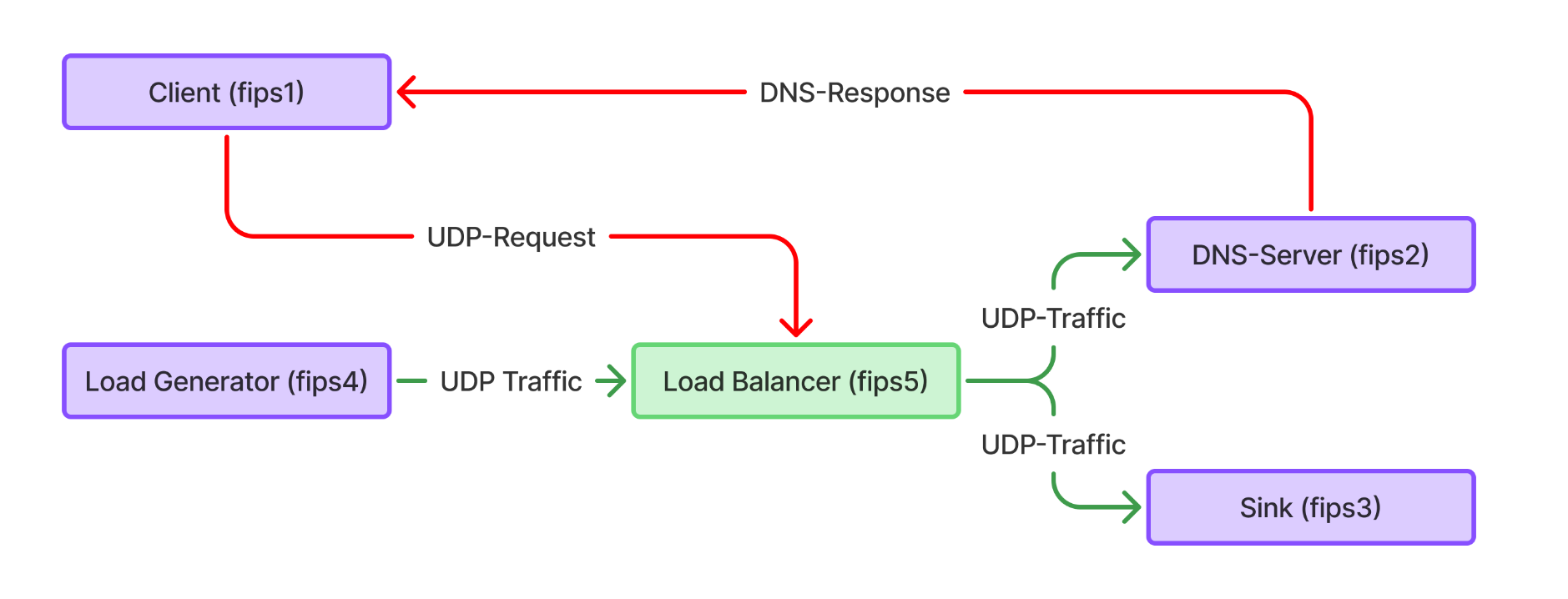}
    \caption{Load Balancer Setup}%
    \label{fig:measurement-setup}
  \end{center}
\end{figure*}

From these metrics, we calculated the actual processing bandwidth by 
\[
\text{Bandwidth (in bit/s)} = \text{pps} * \text{data in bytes} * 8
\]
In our experiments, we observed the behaviour that, even when we applied the
rules for a simple input-output forwarding on the
Bluefield, the traffic was extremely limited to about 100,000 packets per second
with a bandwidth of about 1 Gbit/s. Additionally, we noticed that the packets were routed to
an ARM Core of the Bluefield CPU\@. This was unexpected since it was never
declared to handle packets on the CPU\@. 
After a deeper investigation, it turned
out that Open vSwitch needed to be configured so that the traffic on
the physical function (pf1hpf) needed to be manually routed to port-1. This was done by using the
ovs-ctl cli tool with the following parameters: 

\begin{minted}{bash}
ovs-ofctl -O OpenFlow12 add-flow ovsbr2 \ 
  ip,in_port=pf1hpf,actions=output:p1
\end{minted}

\section{Evaluation}\label{sec:evaluation}

We conducted three different experiments to evaluate the performance of the
Bluefield Architecture with a simple DNS load balancing server. The core idea was
to set up a high traffic DNS service where the ingress needs to be equally
distributed among a defined set of backends with DNS servers running.

\subsection{Description of Testbed}
In order to make sure that the tests are not performance-limited by the actual
implementation of a DNS server, we decided to create our own DNS server with a
reduced feature set. 
A detailed overview of our testbed can be seen in
Figure~\ref{fig:measurement-setup}. The nodes fips2 and
fips3 were used as backends. Fips5 is the load balancer equipped with the
Bluefield-3. Requests were generated on fips4,
which functions as the load generator running T-Rex~\cite{trex}. For the latency
experiments, another client (fips1) was used to send and receive the DNS requests so
that the actual latency could be measured without any interference from other
load on the CPU\@. The hardware specifications of fips1--4 can be seen in
Table~\ref{tab:fips-compute-components}, and the specifications of fips5
and the Bluefield-3 were previously described in
Table~\ref{tab:bf3-compute-components}.

The hosts are connected with a 100 Gbit/s Nvidia MSN2100-CB2F switch.

All measurements were repeated at least three times, and the median values are shown
in the following figures. The observed variation coefficient is below
 0.08\% for all measurements except Figure~\ref{fig:rtt}. For all measurements,
 we also verified that the requested load of
the traffic generator was actually achieved using the interface's hardware
counters\footnote{\texttt{ethtool} was used to query hardware counters before and after the
experiment.}.

\begin{table}[t]
  \centering
  \caption{Compute Components of Servers fips1--4.}%
  \label{tab:fips-compute-components}
  \begin{tabularx}{.5\textwidth}{lR}
    \toprule
    Resource   & Host \\
    \midrule
    CPU        &  Intel Xeon Silver 4314 \\
    Cores      &  16  \\
    Threads    &  32  \\
    Frequency  & 2.4~GHz        \\
    RAM        & 128~GB         \\
    L1D Cache  & 16 $\times$ 48~KB  \\
    L1I Cache  & 16 $\times$ 32~KB  \\
    L2 Cache   & 16 $\times$ 1.25~MB   \\
    L3 Cache   & 24~MB  \\
    \bottomrule
  \end{tabularx}
\end{table}

\subsection{RQ2: Do DOCA Flow applications process with zero overhead, i.e.\ do no packets get dropped below
the maximum throughput? 
}
\begin{figure}[h!]
    \includegraphics[width=0.5\textwidth]{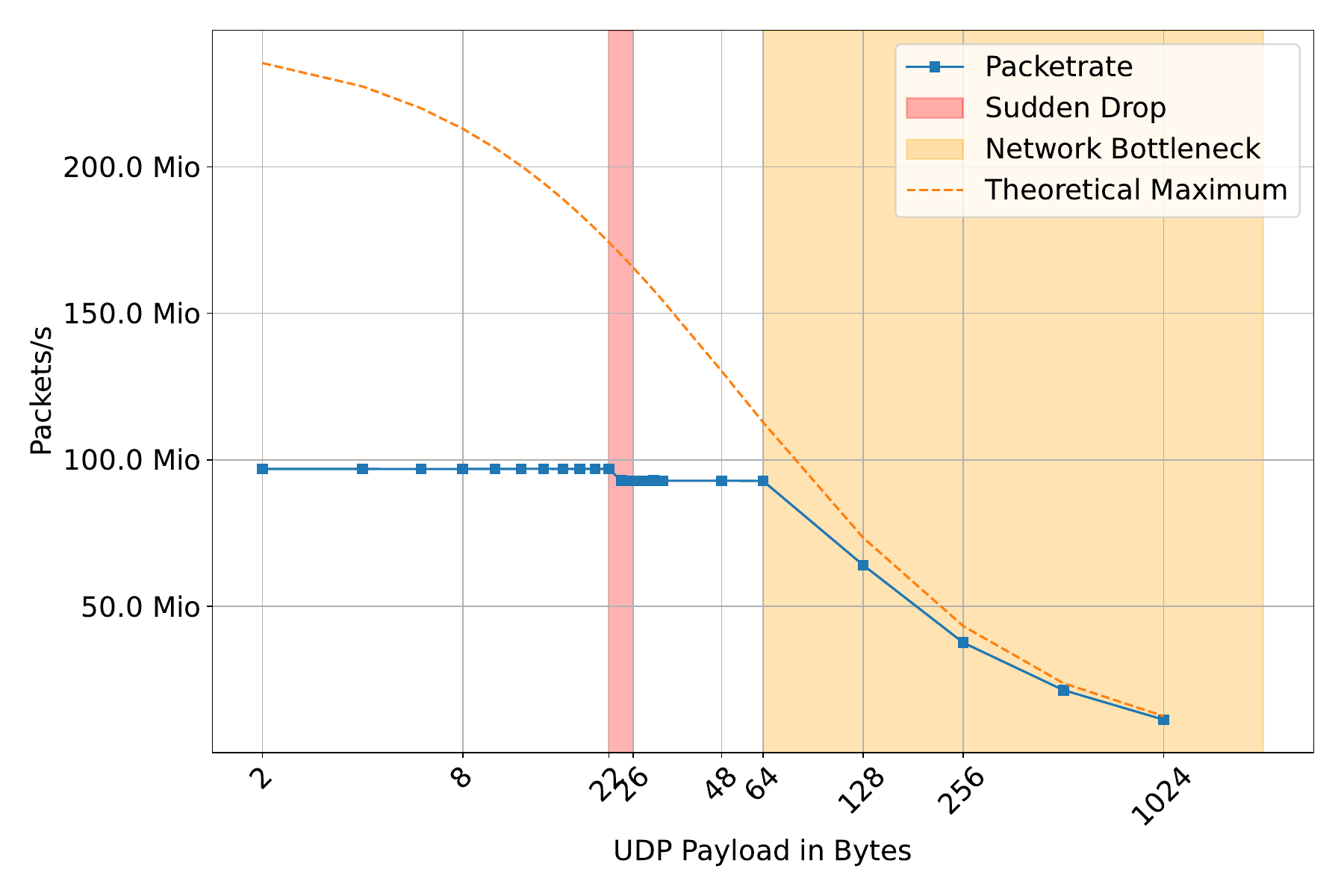}
    \caption{Throughput (pps) Measurements of XenoFlow}%
    \label{fig:xenoflow-pps}
\end{figure}
\begin{figure}[h!]
    \includegraphics[width=0.5\textwidth]{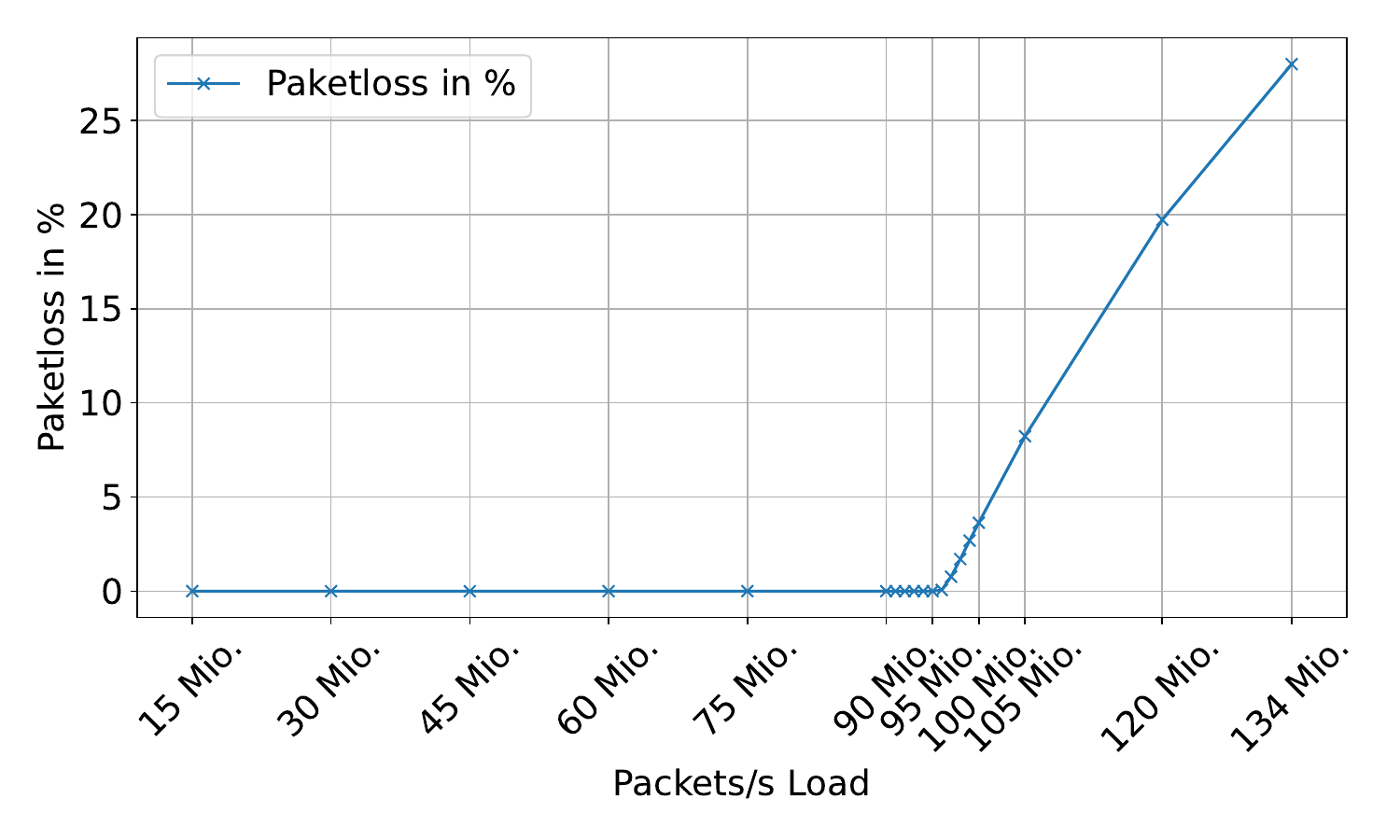}
    \caption{Paketloss Measurements of XenoFlow at UDP Payload Size of 22
    Bytes.}%
    \label{fig:xenoflow-paketloss}
\end{figure}
As seen in Figure~\ref{fig:xenoflow-pps}, the maximum  packets per
second of the Bluefield-3 is around 96.7 million for small packet sizes. 
It shall be noted that at a UDP payload size of exactly and above 23 bytes, the
maximum throughput suddenly drops to 92.8 million packets per second. 23 bytes UDP
Payload is equivalent to 65 bytes Ethernet packet size w/o frame checksum. This
may be a hint that the Bluefield-3 eSwitch has an internal limitation at
64~bytes.  All traffic that exceeds the possible 96.7 million packets/s for
packets smaller than 23 bytes is
getting dropped, as can be seen in Figure~\ref{fig:xenoflow-paketloss}.
Hence, the answer to RQ2 is: No, the Bluefield-3 drops small packets before the
maximum throughput is reached.

\subsection{RQ3: Can the
BlueField-3 achieve line rate
with a load balancer implemented using the
DOCA Flow API?}
\begin{figure}[h!]
    \includegraphics[width=0.5\textwidth]{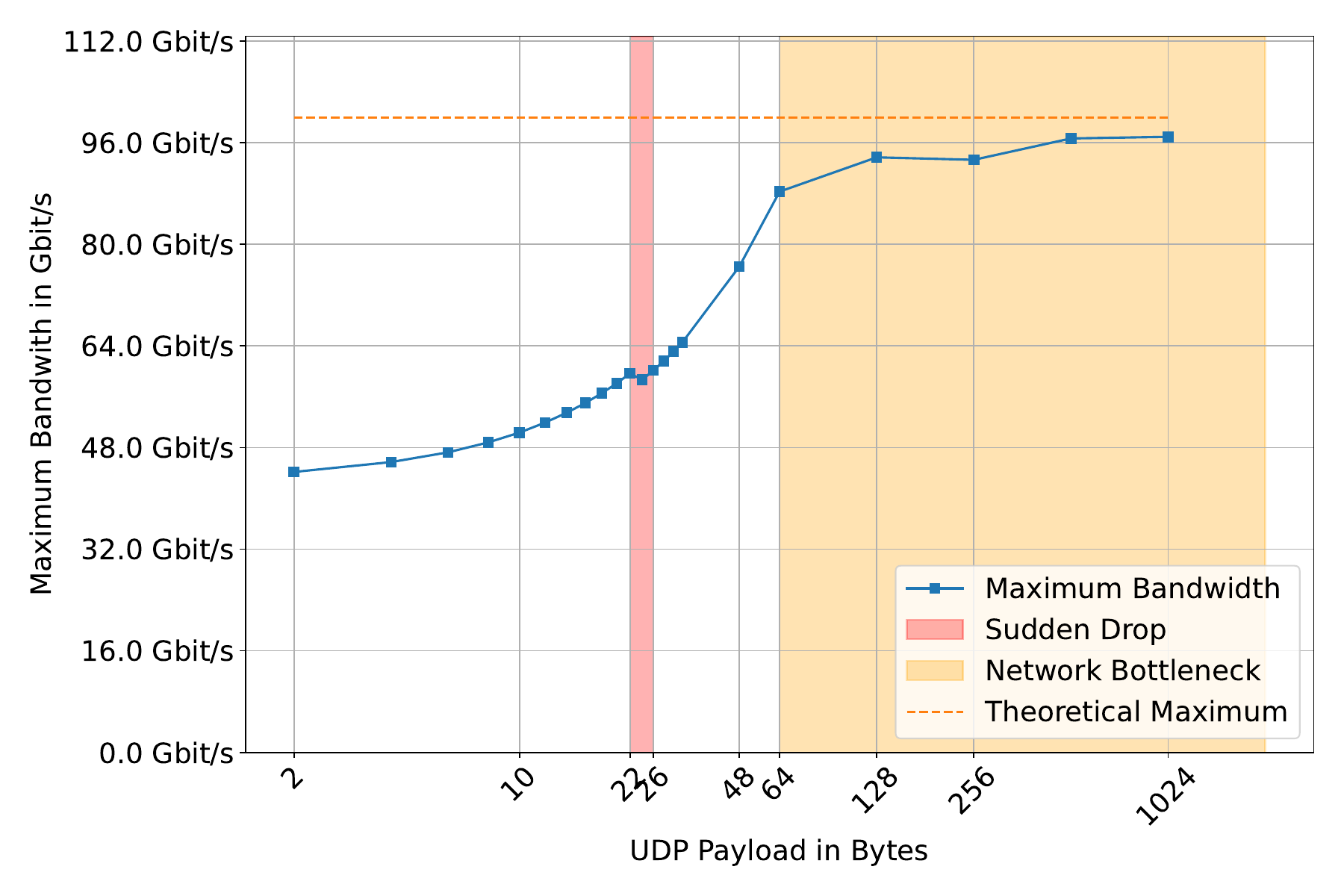}
    \caption{Bandwidth Measurements of XenoFlow}%
    \label{fig:xenoflow-bandwidth}
\end{figure}
As stated by Nvidia, the Bluefield-3 is capable of processing packets at line rate
(400 Gbit/s)~\cite{NV_BF_MARKETING}. 
Even though the Bluefield-3 used for this evaluation has two 100 Gbit/s ports,
the packet processing cores of the accelerated programmable pipeline are, to our
knowledge, the same as in the 400 Gbit/s version. Hence, we expect to achieve 100
Gbit/s. Figure~\ref{fig:xenoflow-bandwidth} illustrates that
XenoFlow barely achieves line rate for UDP Payload beyond 1024~bytes. Given that the current version of
XenoFlow inserts only two entries into the root pipeline, the figure in fact also
quantifies the extent to which the Bluefield-3 falls short of the advertised
line rate (indicated by the dotted red line). 
That said,
the Bluefield 3 is not able to achieve line rate regardless of the actual size of
the packets.

\subsection{RQ4: How does our load balancer affect the
end to end latency?}%
\label{sec:rq4}

\begin{figure}[h!]
\includegraphics[width=0.5\textwidth]{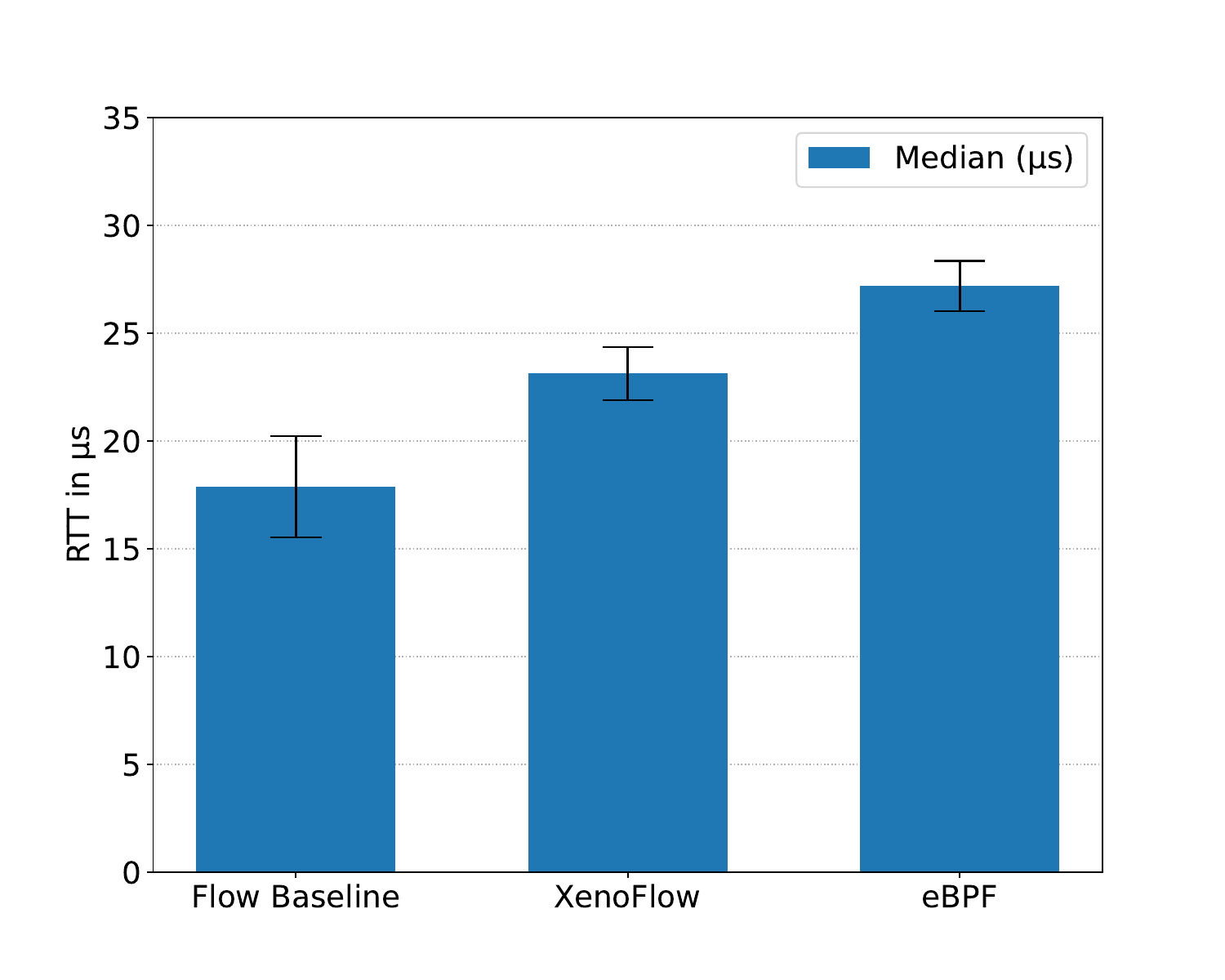}
\caption{Latency Comparison}%
\label{fig:rtt}
\end{figure}

To measure the introduced latency by the load balancer XenoFlow in front of
a service, we measured the round trip time of a DNS request. Since our baseline
measurements without the load balancer and bind9 as DNS server showed a standard
deviation of 35 $\mu s$ a more stable backend with a lower variance and latency
was needed. A DOCA Flow program rewriting the clients requests destination MAC
address was used as backend. The results can be seen in
Figure~\ref{fig:rtt}. We present the median and standard deviation of a total of
2700 measurements. We compared the added latency of
XenoFlow to an equally simple eBPF variant, showing the advantage of hardware
offloading and being closer to the network. While XenoFlow adds only about 5.2 $\mu s$
of latency, the eBPF variant adds about 9.3~$\mu s$ of latency. This is a 44\% latency
reduction.
Compared to a default bind9 installation on Ubuntu 24.04
querying a local zone file with a round trip time of about 102 $\mu s$ the
additional overhead of XenoFlow load balancer is negligible.

\subsection{RQ5: How does load affect the performance of the load balancer?}
\begin{figure}[h!]
\includegraphics[width=0.5\textwidth]{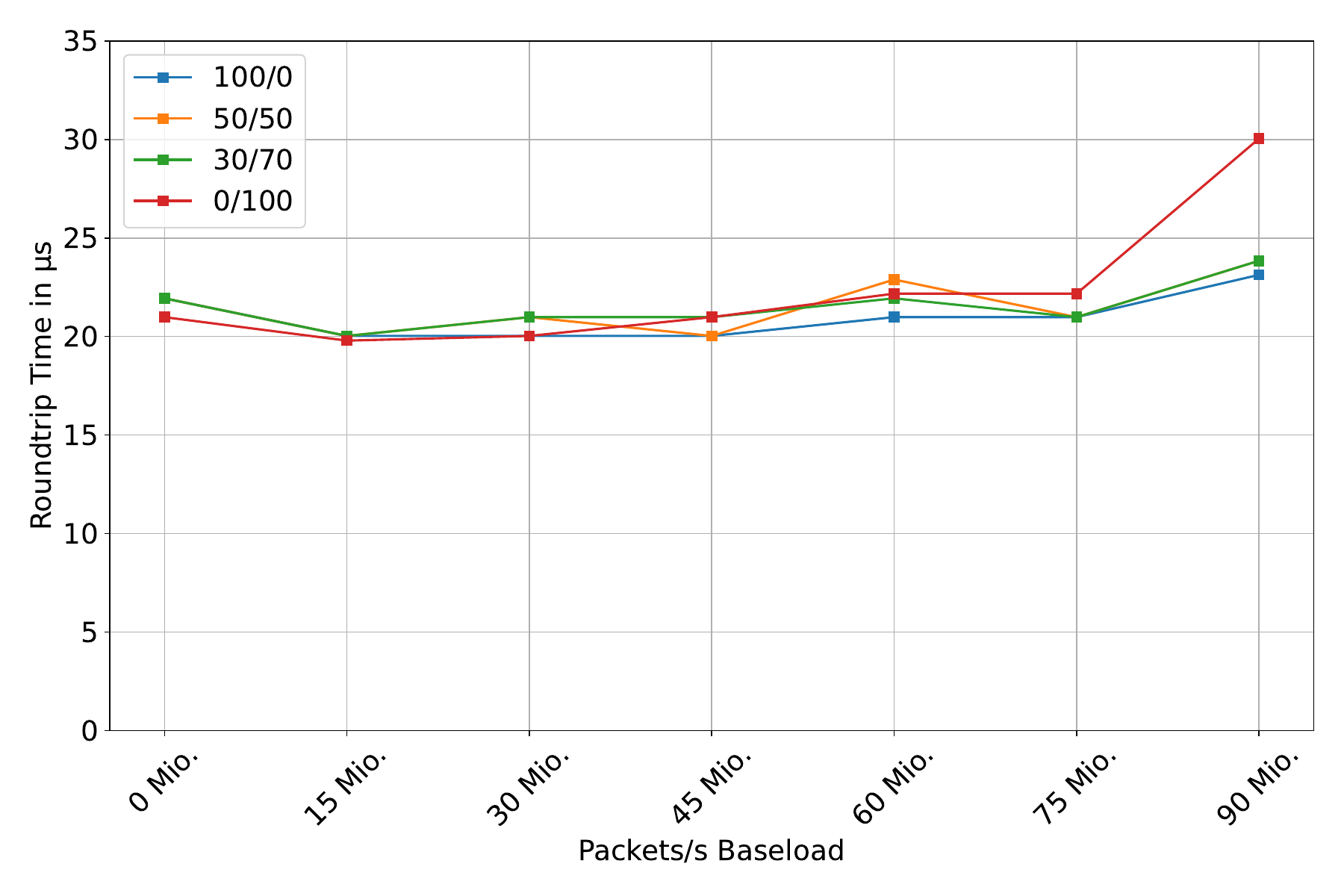}
\caption{Latency under Load}%
\label{fig:xenoflow-load-latency}
\end{figure}

To investigate the effect of load on the latency, we conducted four experiments
with different distributions of the traffic. In these experiments background
load up to 90 Mpps was generated on fips4 using T-Rex. Fips1 was used as a measure 
node where the RTT was recorded. A packet sent by fips1 is distributed to fips2
by the load balancer and then forwarded back to fips1 by the eBPF service
introduced in Section~\ref{sec:rq4} (see Figures~\ref{fig:measurement-setup}).
The results are shown in Figure~\ref{fig:xenoflow-load-latency}. In the single
endpoint experiment we used only one backend (fips2). In the other experiments
two backends (fips2 and fips3) were used. The notation x/y means that x\% of
the traffic was sent to fips2 and y\% to fips3. For all measurements below 90
Mpps the latency stays constant. Only at 90 Mpps an increase in latency for the
experiment where all traffic is sent to fips3 (sink) can be observed. 
However, a slight
increase in latency can be observed for the other distributions, which suggests
that the DOCA Flow API is not
completely zero-overhead as suggested by Nvidia~\cite{NV_BF_MARKETING}. 
Overall, XenoFlow maintains a nearly constant and low-latency profile, even under high load.



\section{Discussion}

In the following, we discuss the research questions RQ1: Can we implement a
simple load balancer using the Bluefield DOCA Flow API\@?

Our prototype XenoFlow clearly shows the potential of the Bluefield-3.
Even though the Bluefield-3 is not able to achieve line rate for small packets,
it still outperforms comparable eBPF implementations on the host.
The intended simplicity of XenoFlow makes it currently incomparable to other
load balancers like e.g.\ Katran. In future work, we plan to extend XenoFlow to
a more feature-complete load balancer.

We successfully implemented a basic L3 load balancer capable of
distributing UDP traffic among a predefined set of backends. The XenoFlow
prototype can easily be extended to more backends by adding more entries with a
specified filter mask and MAC address rewrite action. To address the issue of
uneven
distribution of traffic regarding its IPv4 source addresses, a hash-based
distribution could be implemented by using a hash pipeline with the predefined
hash functions of the DOCA Flow API\@. However, more complex forwarding
strategies would require more complex hash functions like Maglev
hashing~\cite{maglev} or a table storing the current mapping of clients to
backends~\cite{PCC}. Although entries in a pipe could be used to store
such mappings, the limited number of entries (probably 256K) and the slow update
latency (305$\mu s$)~\cite{laconic} may render this approach impracticable.
However, more research is needed to verify this assumption.

The Bluefield-3 also includes the DPA\@. More complex load balancing strategies
could be implemented on the DPA cores to overcome the limitations of the eSwitch.

As an alternative to the DOCA Flow API, the DOCA Pipeline Language
(DPL)~\cite{NV_DPL_HWARC} could be used to implement a load balancer. DPL is a
domain-specific programming language based on the P4-16 language while
introducing Nvidia-specific semantics and architecture tailored for the
BlueField platform. There exists a DPL example that hashes the five-tuple of a packet
and uses the resulting hash as source port~\cite{NV_DPL_EX}. However, a DPL
implementation is restricted to the same capabilities of the eSwitch as the DOCA
Flow API\@. Furthermore, the DPL compiler only supports a subset of the P4-16
language~\cite{NV_DPL_HWARC}.

\section{Conclusion}
In this paper, we presented XenoFlow, a simple L3 load balancer implemented
using the DOCA Flow API on the Bluefield-3 DPU\@. We showed the current
performance capabilities and limitations of the Bluefield-3 and the DOCA Flow
API\@. Our evaluation shows that the Bluefield-3 is not able to achieve line rate
for small packets. However, with a 44\%~lower latency, XenoFlow still outperforms a comparable eBPF
implementation on the host. Furthermore, XenoFlow achieves this low latency even
under high load. The hybrid architecture of the Bluefield-3
makes it a promising platform to integrate more functionality into the DPU, e.g.
an intrusion detection system or a firewall. However, more research is needed to
determine the capabilities of the Bluefield-3 in a real world scenario.

The Nvidia DOCA documentation is sparse and rarely provides substantive
assistance. Moreover, the publicly available documentation repeatedly
references an internal API
specification\footnote{\url{https://docs.nvidia.com/doca/sdk/doca-flow/index.html\#src-3925508478_id-.DOCAFlowv3.1.0-API}},
which is not accessible to external developers. The code-level documentation (restricted to header files, as the implementation is closed source) 
also lacks substance and appears to be largely auto-generated.
Consequently, development that relies on features not covered by the sample
applications is slow and error-prone.

\renewcommand*{\thefootnote}{\fnsymbol{footnote}}
\footnotetext[1]{The source code and all artefacts are available under
\url{https://gitup.uni-potsdam.de/bsvs/public/xenoflow}.}


\bibliographystyle{elsarticle-num}
\biboptions{sort&compress}
\setlength{\bibsep}{0.0pt}
\renewcommand*{\bibfont}{\small}
\bibliography{literature-BF3}

\begin{thebibliography}{10}
\expandafter\ifx\csname url\endcsname\relax
  \def\url#1{\texttt{#1}}\fi
\expandafter\ifx\csname urlprefix\endcsname\relax\def\urlprefix{URL }\fi
\expandafter\ifx\csname href\endcsname\relax
  \def\href#1#2{#2} \def\path#1{#1}\fi

\bibitem{PCC}
{Tom Barbette and Chen Tang and Haoran Yao and Dejan Kosti{\'c} and Gerald Q.
  Maguire Jr. and Panagiotis Papadimitratos and Marco Chiesa},
  \href{https://www.usenix.org/conference/nsdi20/presentation/barbette}{{A
  {High-Speed} {Load-Balancer} Design with Guaranteed
  {Per-Connection-Consistency}}}, in: {17th USENIX Symposium on Networked
  Systems Design and Implementation (NSDI 20)}, USENIX Association, Santa
  Clara, CA, 2020, pp. 667--683.
\newline\urlprefix\url{https://www.usenix.org/conference/nsdi20/presentation/barbette}

\bibitem{HAProxy}
W.~Tarreau.
\newblock \href{https://www.haproxy.org/}{{HAProxy}} [online, cited
  2025-08-21].

\bibitem{IPVS}
{Zhang, Wensong and others}, {Linux Virtual Server for Scalable Network
  Services}, in: {Ottawa Linux Symposium}, Vol. 2000, 2000.

\bibitem{DPDK}
{The Linux Foundation}.
\newblock \href{https://www.dpdk.org/}{{DPDK}} [online, cited 2025-08-21].

\bibitem{DPVS}
{iQiYi QLB}.
\newblock \href{https://github.com/iqiyi/dpvs}{{DPVS}} [online, cited
  2025-08-21].

\bibitem{maglev}
{Eisenbud, Danielle E. and Yi, Cheng and Contavalli, Carlo and Smith, Cody and
  Kononov, Roman and Mann-Hielscher, Eric and Cilingiroglu, Ardas and Cheyney,
  Bin and Shang, Wentao and Hosein, Jinnah Dylan}, {Maglev: A Fast and Reliable
  Software Network Load Balancer}, in: {Proceedings of the 13th Usenix
  Conference on Networked Systems Design and Implementation}, NSDI'16, USENIX
  Association, USA, 2016, p. 523–535.

\bibitem{xdp}
{H\o{}iland-J\o{}rgensen, Toke and Brouer, Jesper Dangaard and Borkmann, Daniel
  and Fastabend, John and Herbert, Tom and Ahern, David and Miller, David},
  {The eXpress Data Path: Fast Programmable Packet Processing in the Operating
  System Kernel}, in: {Proceedings of the 14th International Conference on
  Emerging Networking EXperiments and Technologies}, CoNEXT '18, Association
  for Computing Machinery, New York, NY, USA, 2018, p. 54–66.
\newblock \href {https://doi.org/10.1145/3281411.3281443}
  {\path{doi:10.1145/3281411.3281443}}.

\bibitem{katran}
Facebook.
\newblock \href{https://github.com/facebookincubator/katran}{{Katran}} [online,
  cited 2025-08-21].

\bibitem{SilkRoad}
{Miao, Rui and Zeng, Hongyi and Kim, Changhoon and Lee, Jeongkeun and Yu,
  Minlan}, {SilkRoad: Making Stateful Layer-4 Load Balancing Fast and Cheap
  Using Switching ASICs}, in: {Proceedings of the Conference of the ACM Special
  Interest Group on Data Communication}, SIGCOMM '17, Association for Computing
  Machinery, New York, NY, USA, 2017, p. 15–28.
\newblock \href {https://doi.org/10.1145/3098822.3098824}
  {\path{doi:10.1145/3098822.3098824}}.

\bibitem{LBSNIC21}
{Cui, Tianyi and Zhang, Wei and Zhang, Kaiyuan and Krishnamurthy, Arvind},
  {Offloading Load Balancers onto SmartNICs}, in: {Proceedings of the 12th ACM
  SIGOPS Asia-Pacific Workshop on Systems}, APSys '21, ACM, New York, NY, USA,
  2021, p. 56–62.
\newblock \href {https://doi.org/10.1145/3476886.3477505}
  {\path{doi:10.1145/3476886.3477505}}.

\bibitem{COMPBF}
{Michalowicz, Benjamin and Suresh, Kaushik Kandadi and Subramoni, Hari and
  Panda, Dhabaleswar K. DK and Poole, Steve}, {Battle of the BlueFields: An
  In-Depth Comparison of the BlueField-2 and BlueField-3 SmartNICs}, in: {2023
  IEEE Symposium on High-Performance Interconnects (HOTI)}, 2023, pp. 41--48.
\newblock \href {https://doi.org/10.1109/HOTI59126.2023.00020}
  {\path{doi:10.1109/HOTI59126.2023.00020}}.

\bibitem{laconic}
{Tianyi Cui and Chenxingyu Zhao and Wei Zhang and Kaiyuan Zhang and Arvind
  Krishnamurthy}, {Laconic: Streamlined Load Balancers for SmartNICs} (2024).
\newblock \href {http://arxiv.org/abs/2403.11411} {\path{arXiv:2403.11411}}.

\bibitem{HC33_NV_MARKETING}
{Idan Burstein}, {Nvidia Data Center Processing Unit {(DPU)} Architecture}, in:
  {IEEE} Hot Chips 33 Symposium, {HCS} 2021, Palo Alto, CA, USA, August 22-24,
  2021, {IEEE}, 2021, pp. 1--20.
\newblock \href {https://doi.org/10.1109/HCS52781.2021.9567066}
  {\path{doi:10.1109/HCS52781.2021.9567066}}.

\bibitem{NV_BF_MARKETING}
Nvidia.
\newblock
  \href{https://www.nvidia.com/en-us/networking/products/data-processing-unit/}{{Nvidia
  Portfolio DPUs}} [online, cited 2025-08-21].

\bibitem{DEMYSTIFYINGBF3}
{Chen, Xuzheng and Zhang, Jie and Fu, Ting and Shen, Yifan and Ma, Shu and
  Qian, Kun and Zhu, Lingjun and Shi, Chao and Zhang, Yin and Liu, Ming and
  Wang, Zeke}, {Demystifying Datapath Accelerator Enhanced Off-path SmartNIC},
  in: {2024 IEEE 32nd International Conference on Network Protocols (ICNP)},
  2024, pp. 1--12.
\newblock \href {https://doi.org/10.1109/ICNP61940.2024.10858560}
  {\path{doi:10.1109/ICNP61940.2024.10858560}}.

\bibitem{PERFBF2}
{Jianshen Liu and Carlos Maltzahn and Craig D. Ulmer and Matthew Leon Curry},
  {Performance Characteristics of the BlueField-2 SmartNIC}, CoRR
  abs/2105.06619 (2021).
\newblock \href {http://arxiv.org/abs/2105.06619} {\path{arXiv:2105.06619}}.

\bibitem{laconic-early}
T.~Cui, W.~Zhang, K.~Zhang, A.~Krishnamurthy, Offloading load balancers onto
  smartnics, in: Proceedings of the 12th ACM SIGOPS Asia-Pacific Workshop on
  Systems, APSys '21, Association for Computing Machinery, New York, NY, USA,
  2021, p. 56–62.
\newblock \href {https://doi.org/10.1145/3476886.3477505}
  {\path{doi:10.1145/3476886.3477505}}.

\bibitem{NV_BF_HWARC}
Nvidia.
\newblock
  \href{https://docs.nvidia.com/doca/sdk/hardware+architecture/index.html}{{DOCA
  Documentation Hardware Architecture}} [online, cited 2025-08-21].

\bibitem{INTEL_ESWITCH}
Intel.
\newblock
  \href{https://edc.intel.com/content/www/ca/fr/design/products/ethernet/appnote-e810-eswitch-switchdev-mode-config-guide/eswitch-mode-switchdev-and-legacy/}{{eSwitch
  Mode (Switchdev and Legacy)}} [online, cited 2025-08-21].

\bibitem{MELLANOX_ASAP}
Nvidia.
\newblock
  \href{https://network.nvidia.com/files/doc-2020/sb-asap2.pdf}{{Mellanox
  ASAP}} [online, cited 2025-08-21].

\bibitem{NV_APAP_LIM}
Nvidia.
\newblock
  \href{https://docs.nvidia.com/networking/display/mlnxenv23102131201lts/ovs+offload+using+asap%C2%B2+direct}{{Documentation
  OVS Offload Using ASAP Direct}} [online, cited 2025-08-21].

\bibitem{NV_BF_PIPE}
Nvidia.
\newblock
  \href{https://docs.nvidia.com/doca/archive/doca-v1.5.2/flow-programming-guide/index.html#pipe-mode}{{DOCA
  Documentation Pipe Mode}} [online, cited 2025-08-21].

\bibitem{NV_FLOW_EX}
Nvidia.
\newblock
  \href{https://docs.nvidia.com/doca/archive/doca-v1.5.1/flow-samples/index.html#flow-hairpin}{{DOCA
  Documentation Flow Samples}} [online, cited 2025-08-21].

\bibitem{trex}
C.~systems.
\newblock \href{https://trex-tgn.cisco.com/trex/doc/}{{TRex realistic traffic
  generator}} [online, cited 2025-08-21].

\bibitem{NV_DPL_HWARC}
Nvidia.
\newblock
  \href{https://docs.nvidia.com/doca/sdk/dpl+system+overview/index.html}{{DOCA
  Documentation DPL System Overview}} [online, cited 2025-08-21].

\bibitem{NV_DPL_EX}
Nvidia.
\newblock
  \href{https://docs.nvidia.com/doca/sdk/entropy+in+hash-based+load+balancing+example/index.html}{{Entropy
  in Hash-Based Load Balancing Example}} [online, cited 2025-08-21].

\end{thebibliography}

\end{document}